\documentclass[namedreferences]{SolarPhysics}
\usepackage[optionalrh]{spr-sola-addons} 

\usepackage{graphicx}        
\usepackage{amssymb}        
\usepackage{color}           
\usepackage{url}             

\newcommand{\cS}{{\cal S}}
\newcommand{\cN}{{\cal N}}

\newcommand{\ii}{{\rm i}}
\newcommand{\?}{_{\shortmid}^{\shortmid}}

\begin{document}

\begin{article}

\begin{opening}
\title{Fourier Analysis of Gapped Time Series: Improved Estimates of Solar and Stellar Oscillation Parameters}
\author{Thorsten \surname{Stahn} \sep Laurent \surname{Gizon}}
\runningauthor{T. Stahn, L. Gizon}
\runningtitle{Improved Estimates of Solar and Stellar Oscillation Parameters}
\institute{Max-Planck-Institut f\"ur Sonnensystemforschung, 37191 Katlenburg-Lindau, Germany, e-mail:\url{gizon@mps.mpg.de}}
\date{Received 2 January 2008 / Accepted 15 March 2008}
\begin{abstract}
Quantitative helio- and asteroseismology require very precise measurements of the frequencies, amplitudes, and lifetimes of the global modes of stellar oscillation. It is common knowledge that the precision of these measurements depends on the total length ($T$), quality, and completeness of the observations. Except in a few simple cases, the effect of gaps in the data on measurement precision is poorly understood, in particular in Fourier space where the convolution of the observable with the observation window introduces correlations between different frequencies. Here we describe and implement a rather general method to retrieve maximum likelihood estimates of the oscillation parameters, taking into account the proper statistics of the observations. Our fitting method applies in complex Fourier space and exploits the phase information. We consider both solar-like stochastic oscillations and long-lived harmonic oscillations, plus random noise. Using numerical simulations, we demonstrate the existence of cases for which our improved fitting method is less biased and has a greater precision than when the frequency correlations are ignored. This is especially true of low signal-to-noise solar-like oscillations. For example, we discuss a case where the precision on the mode frequency estimate is increased by a factor of five, for a duty cycle of 15\%. In the case of long-lived sinusoidal oscillations, a proper treatment of the frequency correlations does not provide any significant improvement; nevertheless we confirm that the mode frequency can be measured from gapped data at a much better precision than the $1/T$ Rayleigh resolution. 
\end{abstract}
\end{opening}

\section{Introduction}
Solar and stellar oscillations are a powerful tool to probe the interior of stars. In this paper we classify stellar oscillations into solar-like or deterministic. Solar-like oscillations are stochastically excited by turbulent convection and are present in the Sun and other main-sequence, subgiant, and giant stars (see {\it e.g.}~\citeauthor{Bedding2007}, \citeyear{Bedding2007} and references therein). Deterministic oscillations are seen in classical pulsators and have mode lifetimes much longer than any typical observational run; one of the best studied objects in this class is the pre-white dwarf PG1159$-$035 also known as GW~Vir \cite{Winget1991}.  In practice, observations of solar-like or deterministic pulsations always have an additional stochastic component due to instrumental, atmospheric, stellar, or photon noise. 

An important aspect of helio- and asteroseismology is the determination of the parameters of the global modes of oscillation, especially the mode frequencies. In the case of the Sun, it is known \cite{Woodard1984} that the measurement precision is limited by the stochastic nature of the oscillations (realization noise). \inlinecite{Libbrecht1992} and \inlinecite{Toutain1994} have shown that realization noise is expected to scale like $1/\sqrt{T}$, where $T$ is the total duration of the observation. A common practice is to extract the solar mode parameters from the power spectrum using maximum likelihood estimation (MLE, see {\it e.g.}~\opencite{Anderson1990}; \opencite{Schou1992thesis}; \opencite{Toutain1994}; \opencite{Appourchaux1998}; \opencite{Appourchaux2000}).  In its current form, however, this method of analysis is only valid for uninterrupted time-series. This is a significant limitation because gaps in the data are not uncommon (daily cycle, bad weather, technical problems).  The gaps complicate the analysis in Fourier space: the convolution of the data with the observation window leads to correlations between the different Fourier components.  The goal of this paper is to extend the Fourier analysis of solar and stellar oscillations to time series with gaps, using appropriate maximum likelihood estimators based on the correct statistics of the data.

Section~\ref{section:problem} poses the problem of the analysis of gapped time series in Fourier space. In Section~\ref{section:pdf_correlation} we derive an expression for the joint probability density function (PDF) of the observations, taking into account the frequency correlations. Our answer is consistent with an earlier (independent) derivation by \inlinecite{Gabriel1994}. Based on this PDF, we derive maximum likelihood estimators in Section~\ref{section:likelihood_estimation}. In Section~\ref{section:fitting_nocorr} we recall the ``old method'' of maximum likelihood estimation based on the unjustified assumption that frequency bins are statistically independent. Section~\ref{section:test_setup} explains the set up of the Monte-Carlo simulations, used to test the fitting methods on artificial data sets. In Section~\ref{section:fitting_results} we present the results of the Monte-Carlo simulations and compare the new and old fitting methods.  For the sake of simplicity, we consider only one mode of oscillation at a time (solar-like or sinusoidal). We present several cases for which our new fitting method leads to a significant improvement in the determination of oscillation parameters, and in particular the mode frequency. 

\section{Statement of the Problem}
\label{section:problem}

\subsection{The Observed Signal in Fourier Space}
\label{section:convolution}
Let us denote by $\tilde{y}=\{\tilde{y}_i\}$ the time series that we wish to analyse.  It is sampled at times $t_i=i \Delta t$, where $i$ is an integer in the range $0\le i\le N-1$, and $\Delta t=$~one minute is the sampling time. 
All quantities with a tilde are defined in the time domain. The total duration of the time series is $T=N\Delta t$. By choice, all of the missing data points were assigned the value zero: this enables us to work on a regularly sampled time grid.   Formally, we write
\begin{equation}
  \tilde{y}_i = \tilde{w}_i \tilde{x}_i , \quad i = 0, 1, \dots, N-1.
  \label{equ:obs_signal}
\end{equation}
where  $\tilde{x}$ is the uninterrupted time series that we {\it would} have observed if there had been no gaps and $\tilde{w}$ is the window function defined by $\tilde{w}_i=1$ if an observation is recorded at time $t_i$ and $\tilde{w}_i=0$ otherwise. The $\tilde{x}$ is drawn from a random process, whose statistical properties will be discussed later.

We define the discrete Fourier transform $\hat{y}$ of $\tilde{y}$ by
\begin{equation}
  \hat{y}_j=\frac{1}{N} \sum_{i=0}^{N-1} \tilde{y}_i \; {\rm e}^{- \ii 2\pi \nu_j t_i}  \quad {\rm for} \; j \in \mathbb{N} ,
  \label{equ:FT}
\end{equation}
where $\nu_j = j \Delta \nu$ is the frequency and $\Delta \nu = 1 / N\Delta t $.  Note that $\hat{y}_j = \hat{y}_{N-j}^*$ and  $\hat{y}_j = \hat{y}_{-j}^*$, where the star denotes the complex conjugate. The Fourier transform has periodicity $1/\Delta t$ or twice the Nyquist frequency.

\begin{figure}[t]
  \centering
  \includegraphics[angle=90,width=0.95\textwidth]{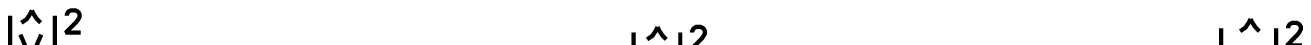}
  \caption{Schematic representation in Fourier space  of the convolution of the signal $\hat{x}$ with the window function $\hat{w}$. For the sake of simplicity, only the power spectra of the different quantities are shown here. Panel (a) shows the window function $\hat{w}$ and its cutoff frequency $\nu_p$, panel (b) the unconvolved signal $\hat{x}$, and panel (c) the observed signal $\hat{y}$. Note that the selected section of the observed signal, starting at frequency $\nu_p$, is of  length $M$, while the unconvolved signal is of length $(M+2p)$. }
 \label{fig:convolution_scheme}
\end{figure}

Our intention is not to fit the complete Fourier spectrum, but a rather small interval that contains one (or a few) modes of stellar oscillation. Thus, we extract a section of the data of length $M$ starting from a particular frequency $\nu_q$, as shown in Figure~\ref{fig:convolution_scheme}(c). This subset of the data is represented by the vector $y=[y_0, y_1, \cdots, y_{M-1}]^T$ with components
\begin{equation}
  y_i = \hat{y}_{q+i}  , \quad i = 0, 1, \dots, M-1.
\end{equation}
Using the above definition of the Fourier transform, the vector $y$ is given by the convolution of $\hat{x}$ with the window $\hat{w}$:
\begin{equation}
  y_i = \sum_{j=-p}^{M+p-1} \hat{w}_{i-j} \; \hat{x}_{q+j}.
  \label{equ:fourier_elements}
\end{equation}
The integer $p$ in Equation (\ref{equ:fourier_elements}) refers to the cutoff frequency $\nu_p$ beyond which the observation window has no significant power. Truncating the window function at frequency $\nu_p$ is a simplification of the general problem. Our main goal, however, is, given a known window function, to study its effects on the determination of the parameters of stellar oscillations. Figure \ref{fig:convolution_scheme} is a schematic representation in Fourier space of the convolution of a single mode of oscillation by the window function.  The observed signal is spread over some frequency range and, as we shall see later,  its statistical properties are affected.

We note that, in practice, one can never completely isolate one single mode of oscillation in the power spectrum. In particular, other modes with frequencies outside of the fitting range can leak into it after convolution by the temporal window function. Hence, fitting one mode of oscillation is a simplification. But our first objective is to try to study the effects of gaps, independently from the complications associated with a badly specified model. 

Equation (\ref{equ:fourier_elements}) can be rewritten in matrix form as 
\begin{equation}
  y =W x ,
  \label{equ:fourier_signal}
\end{equation}
where  the vector $x=[x_0, x_1, \dots, x_{M+2p-1}]^T$ of length $M+2p$ is defined by
\begin{equation}
x_i = \hat{x}_{q-p+i}  , \quad i = 0, 1, \dots, M + 2p - 1 ,
\end{equation}
and $W=[W_{ij}]$ is the $M\times(M+2p)$ rectangular window matrix with elements $W_{ij} = \hat{w}_{i-j+p}$, where $i = 0,1, \dots, M-1$ and $j = 0,1, \dots, M+2p-1$:
\begin{equation}
W = \left[\begin{array}{*{9}{c}}
    \hat{w}_p & \dots & \hat{w}_0 & \dots & \hat{w}_{-p} & & & & \\
    & \ddots & & \ddots & & \ddots & & 0 & \\
    & & \hat{w}_p & \dots & \hat{w}_0 & \dots & \hat{w}_{-p} & & \\
    & 0 & & \ddots & & \ddots & & \ddots & \\
    & & & & \hat{w}_p & \dots & \hat{w}_0 & \dots & \hat{w}_{-p} \\
  \end{array} \right].
\label{equ:windowmatrix}
\end{equation}
Note that $\hat{w}_i=\hat{w}^*_{-i}$ and that $W$ is of rank $M$.

Equation~(\ref{equ:fourier_signal}) is the master equation. Our goal is to extract the stellar oscillation parameters (contained in $x$), given the uncomplete information $y$.

\subsection{Statistics of the Unconvolved Signal}
\label{section:signal_stats}
 
Here we describe the basic assumptions that we make about the statistics of the data in the Fourier domain.
The unconvolved signal [$x$] consists of a deterministic component [$d$] and a zero-mean stochastic component [$e$] such that
\begin{equation}
x = d + e.  
\label{equ:signal}
\end{equation}
The deterministic component $d$ may include deterministic stellar oscillations that are long-lived compared to the total length of the observation. The stochastic component $e$ may include various sources of noise ({\it e.g.} stellar convection, photon noise, atmospheric noise, {\it etc.}) and stochastically excited pulsations as observed on the Sun.

We assume that the $e_i$ are $M+2p$ independent random variables in the Fourier domain. This is equivalent to saying that the stochastic component of the signal in the time domain is stationary.  We further assume that $e$ is a Gaussian random vector with independent real and imaginary parts and covariance matrix
\begin{equation}
E[ e_i^* e_j ]=\sigma_i^2 \delta_{ij} , \quad i,j = 0,1,\dots,M+2p-1 ,
\label{equ:sigma^2} 
\end{equation}
where $E$ denotes the expectation value and $\sigma_i$ is the standard deviation of $e_i$ at frequency $\nu_i$. One may invoque the central limit theorem to justify the choice of Gaussian distributions. The quantity $\sigma^2_i$ is the expected power spectrum at frequency $\nu_i$, which may include background noise and peaks corresponding to the modes of oscillations \cite{Duvall1986, Appourchaux1998}.  In terms of a complex Gaussian random vector $g$ with unit covariance matrix, $E[ g^* g^T ]=I_{M+2p}$, we can rewrite $e$ as
\begin{equation}
  e = S g,
\end{equation}
where $S$ is the $(M+2p)\times (M+2p)$ diagonal matrix 
\begin{equation}
S={\rm diag}(\sigma_{0}, \sigma_{1}, \dots, \sigma_{M+2p-1}).
\label{equation:matrix_d}
\end{equation}
We emphasize that, although the $e_i$ are uncorrelated random variables, the $y_i$ are correlated because of the multiplication of $x$ by the window matrix [Equation~(\ref{equ:fourier_signal})].

\section{Joint PDF of the Complex Fourier Spectrum}
\label{section:pdf_correlation}

In this section we derive an expression for the joint probability density function of the observed signal $y$. This problem had already been solved by \inlinecite{Gabriel1994}. We reach the same conclusion, independently and with more compact notations. We start by rewriting the master equation, Equation (\ref{equ:fourier_signal}),  as
\begin{equation}
  y = W d + C g, 
  \label{equ:fourier_signal_2}
\end{equation}
where 
\begin{equation}
C=WS
\end{equation}
 is a $M\times (M+2p)$ matrix with rank $M$ and singular value decomposition \cite[chapter 7.3]{Horn1985}
\begin{equation}
  C = U \Sigma V^H .
\end{equation}
Here the superscript $H$ denotes the Hermitian conjugate and $U$ and $V$ are unitary matrices of dimensions $M\times M$ and $(M+2p)\times (M+2p)$ respectively,  {\it i.e.} $U^H U=I_M$ and $V^H V=I_{M+2p}$.   The $M\times (M+2p)$  matrix $\Sigma$ can be written as
\begin{equation}
  \Sigma= [\Lambda \, \? \, 0], \quad \Lambda=\mbox{diag}(\lambda_0,\lambda_1, \dots ,\lambda_{M-1}) ,
  \label{equ:svd}
\end{equation}
where $\lambda_0,  \lambda_1, \dots , \lambda_{M-1}$ are the $M$ (positive) singular values of the matrix $C$. Thus, there exists a vector $\xi = V^H g$ such that
\begin{equation}
  y = W d + U[\Lambda \, \? \, 0] \xi .
  \label{equ:fourier_signal_xi} 
\end{equation}
Since $g$ has unit covariance matrix and $V$ is unitary, the vector $\xi$ is a complex Gaussian random vector of size $M+2p$ with unit covariance matrix. It is obvious from Equation~(\ref{equ:fourier_signal_xi}) that there exists a lower-rank complex Gaussian random vector of length $M$, 
$\eta =[\xi_0 , \xi_1 , \dots , \xi_{M-1}]^T$, such that 
\begin{equation}
  y = W d + U \Lambda \eta.
  \label{equ:signalll}
\end{equation}
The variables $\xi_{M}, \xi_{M+1}, \dots , \xi_{M+2p-1}$ are dummy variables, which do not enter in the description of $y$.  Equation~(\ref{equ:signalll}) is an important step, as the vector $y$ of length $M$ is now expressed in terms of $M$ independent complex Gaussian variables. This enables us to write the PDF of $y$ as
\begin{equation}
  p_{y}(y) = \frac{1}{J} \;  p_{\eta}\left( (U\Lambda)^{-1}(y-W d) \right),
  \label{equ:pdf_transorm}
\end{equation}
where $p_{\eta}(\eta)$ denotes the PDF of $\eta$ and $J$ is the Jacobian of the linear transformation $\eta \rightarrow y$. Since $\eta$ is a complex Gaussian random vector with unit covariance, {\it i.e.} $E[\eta^* \eta^T]=I_M$, we have
\begin{equation}
  p_{\eta}(\eta)=\frac{\exp (-\|\eta\|^2)}{\pi^M} ,
\label{eq.pdf_eta}
\end{equation}
where we used the notation $\|\eta\|^2=\eta^H \eta$. Since $U$ is unitary and $\Lambda$ is diagonal and real, the Jacobian of the transformation is given by
\begin{equation}
  J = |\det(U\Lambda)|^2 = (\det \Lambda)^2  = \prod_{i=0}^{M-1} \lambda_i^2 .
  \label{equ:jacobian}
\end{equation}
Combining Equations~(\ref{equ:pdf_transorm}), (\ref{eq.pdf_eta}), and (\ref{equ:jacobian}), we get the joint PDF of the observed vector $y$:
\begin{equation}
  p_{y}(y) =\frac{\exp (-\| \Lambda^{-1} U^H (y-W d) \|^2)}{\pi^M (\det \Lambda)^2} .
\end{equation}
The above expression is, perhaps, more elegantly written as
\begin{equation}
  p_{y}(y) 
  = \frac{\exp (-\| C^\dagger ({y}-W d) \|^2)}{\pi^M (\det \Lambda)^2}
  \label{equ:pdf} 
\end{equation}
in terms of $C^\dagger$, the $(M+2p) \times M$ Moore-Penrose generalized inverse of $C$ \cite[chapter 7.3]{Horn1985},\begin{equation}
  C^{\dagger} = V \Sigma^\dagger U^H = C^H(CC^H)^{-1} ,
\label{equ:pseudo_inverse}
\end{equation}
where $\Sigma^\dagger$ is the transpose of $\Sigma$ in which the singular values are replaced by their inverse.  One may ask, after the fact, if the quantity $(U\Lambda)^{-1}$ in Equation (\ref{equ:pdf_transorm}) is always defined. The answer would appear to be yes  since the Moore-Penrose generalised inverse of $C$ is perfectly well defined. It is not excluded, however, that some singular values $\lambda_i$ could be infinitesimally small. We have not encountered any such difficulty with the test cases given in Section \ref{section:fitting_results}. Should  $C$ be ill-conditioned in other cases, a simple truncated SVD would help avoiding a numerical problem.

Before discussing the implementation of the method in Section~\ref{section:likelihood_estimation}, we should like to draw attention to a parallel between fitting data with temporal gaps and fitting data with spatial gaps. In order to understand this analogy, we refer the reader to the work of  \citeauthor{Appourchaux1998} (1998, Section 3.3.4) who discuss how to interpret the spatial leaks of non-radial oscillations that arise from the fact that only half of the solar disk can be observed from Earth. Their approach is similar to the one developed in this paper.

\section{Maximum Likelihood Estimation of Stellar Oscillation Parameters}
\label{section:likelihood_estimation}

Let us assume that the stellar oscillation model that we are trying to fit to the data depends on a set of $k$ parameters $\mu=(\mu_0 , \mu_1 , \dots , \mu_{k-1})$. These parameters may be the amplitude, the phase, the frequency, the line asymmetry, the noise level, {\it etc}.  The basic idea of maximum likelihood estimation is to pick the estimate ${\mu_\star}$ that maximizes the likelihood function. The likelihood function is another name for the joint PDF [Equation~(\ref{equ:pdf})] evaluated for the sample data.  In practice, one minimizes 
\begin{equation}
  {\cal L}(\mu) = -\ln p_y = \|C^\dagger (y-W d)\|^2 + 2 \sum_{i=0}^{M-1} \ln \lambda_i + {\rm constant},
  \label{equ:likelihood}
\end{equation}
rather than maximizing the likelihood function itself.
In the above expression, the quantities $C^\dagger$ and $\lambda_i$ all depend implicitly on the model parameters $\mu$ through the covariance matrix  $S$. The vector $d$ also depends on the model parameters in the case deterministic oscillations.  The probability of observing the sample data is greatest if the unknown parameters are equal to their maximum likelihood estimates $\mu_\star$:
\begin{equation}
  \mu_\star = \arg \displaystyle\min_{\mu} \; {\cal L}({\mu}) .
\label{equ:likelihood-estimator}
\end{equation}
The method of maximum likelihood has many good properties \cite{Brandt1970}. In particular, in the limit of a large sample size ($M$ large), the maximum likelihood estimator is unbiased and has minimum variance. 

What is particularly new about our work is the minimization of the likelihood function given by Equation~(\ref{equ:likelihood}). We use the direction set method, or Powell's algorithm, to solve  the minimization problem with a computer. In practice, the result of the fit depends on the initial guess and the fractional tolerance of the minimisation procedure (the relative decrease of ${\cal L}$ in one iteration). The dependence of the fitted parameters on the initial guess is due to the fact that the function ${\cal L}$ may have local minima in addition to the global minimum. We will address this issue in more detail in Section \ref{section:fitting_results}.

\subsection{Special Case: Solar-Like Oscillations}
In the case of solar-like oscillations, there is no deterministic component and the log-likelihood becomes
\begin{equation}
  {\cal L}(\mu) = \|C^\dagger y \|^2 + 2 \sum_{i=0}^{M-1} \ln \lambda_i + {\rm constant}.
\end{equation}

\subsection{Special Case: Deterministic Oscillations plus White Noise}
If background white noise is the only stochastic component then
\begin{equation}
\sigma_i = \sigma_0 = {\rm constant}  , \quad i = 0,1,\dots,M+2p-1 .
\end{equation}
The log-likelihood function becomes
\begin{equation}
  {\cal L}(\mu) =  \frac{1}{\sigma_0^2} \; \|W^\dagger (y-Wd) \|^2 +  M \;  \ln \sigma^2_0  + {\rm constant}.
\end{equation}
Splitting the unknowns $\mu=(\check{\mu}, \sigma_0)$ into the parameters describing the oscillations, $\check{\mu}=(\mu_0 , \mu_1 , \dots , \mu_{k-2})$ and the noise level $\sigma_0$, the minimization problem reduces to finding the most likely estimates
\begin{equation}
  \check{\mu}_\star = \arg \displaystyle \min_{\check{\mu}} \; \|W^\dagger (y-Wd) \|^2 ,
\end{equation}
where $d=d(\check{\mu})$. The noise level is explicitly given by
\begin{equation}
  \sigma_{0 \star} = M^{-1/2} \|W^\dagger [y-Wd(\check{\mu}_\star)] \| .
\end{equation}

\section{The Old Way: Fitting the Power Spectrum Ignoring the Correlations}
\label{section:fitting_nocorr}

Maximum likelihood estimation has been used in the past to infer solar and stellar oscillation parameters, even in the case of gapped time series. The joint PDF of the observations was assumed to be the product of the PDFs of the individual  $y_i$, as if the frequency bins were uncorrelated. For comparison purposes, we briefly review this (unjustified) approach.

According to Equation~(\ref{equ:fourier_signal_2}), the PDF of $y_i$ is a normal distribution
\begin{equation}
p_{y_i}(y_i) = \frac{\exp( -| y_i - \overline{y}_i |^2 / v_i )}{\pi  v_i } 
\end{equation}
with mean 
\begin{equation}
\overline{y}_i = \sum_{j=0}^{M+2p-1} W_{ij} d_j 
\end{equation}
and variance
\begin{equation}
v_i  =  \sum_{j=0}^{M+2p-1} |W_{ij}|^2 \sigma_j^2  .
\end{equation}
Under the (wrong) assumption that the $y_i$ are independent random variables, the joint PDF of $y$ becomes
\begin{equation}
p_y^{\rm nc}(y) =  \prod_{i=0}^{M-1}    p_{y_i}(y_i)  ,
\end{equation}
where the superscript  ``nc'' stands for  ``no correlation''. This joint PDF  uses the correct mean ($\overline{y}_i$) and variance ($v_i$) of the data, but it ignores all the non-vanishing cross-terms $E[y_i^* \,  y_j]$.  In other words, the spread of power implied by the convolution with the window is taken care of,  but not the proper statistics.

Under the same simplifying ``no correlation'' assumption, the log-likelihood function is 
\begin{equation}
{\cal L}^{\rm nc}(\mu)  = \sum_{i=0}^{M-1} \frac{| y_i - \overline{y}_i |^2}{v_i}
+ \sum_{i=0}^{M-1} \ln v_i + {\rm constant} ,
\end{equation}
where the $\overline{y}_i$ and $v_i$ are implicit functions of the model parameters $\mu$.   

\subsection{Special Case: Solar-Like Oscillations}
If the signal has no deterministic component ($d=0$), then the power spectrum [$P_i(\mu)=|y_i|^2$] has the expectation value $\overline{P}_i = E[P_i]= v_i$. Thus, in the case of purely solar-like oscillations, we recover the standard expression \cite{Toutain1994}:
\begin{equation}
{\cal L}^{\rm nc}(\mu)  = \sum_{i=0}^{M-1} \left( \frac{P_i}{\overline{P}_i} 
+   \ln \overline{P}_i \right) + {\rm constant}  \quad {\rm when} \;  d=0.
\label{eq.nc}
\end{equation}
While the above expression is perfectly valid for uninterrupted data, it is not justified when gaps are present. 
The parameters $\mu^{\rm nc}_\star$ that minimize ${\cal L}^{\rm nc}(\mu)$ are not optimal, as will be shown later using Monte-Carlo simulations.

\subsection{Special Case: Deterministic Oscillations plus White Noise}
When $\sigma_i=\sigma_0={\rm constant}$,  the ``no-correlation'' log-likelihood function simplifies to 
\begin{equation}
{\cal L}^{\rm nc}(\mu)  = \frac{1}{\sigma_0^2} \; \frac{ \|y - W d\|^2 }{ \sum_{j=-p}^{p} | \hat{w}_j |^2}
+  M \; \ln \sigma^2_0 + {\rm constant}.  
\end{equation}
The minimization problem becomes
\begin{equation}
  \check{\mu}_\star^{\rm nc} = \arg \displaystyle \min_{\check{\mu}} \; \|y-Wd \|^2 ,
\end{equation}
where $d(\check{\mu})$ depends on the oscillation parameters $\check{\mu}=(\mu_0 , \mu_1 , \dots , \mu_{k-2})$. The noise level is explicitly given by
\begin{equation}
  \sigma_{0 \star}^{\rm nc} = \left(M \sum_{j=-p}^{p} | \hat{w}_j |^2 \right)^{-1/2} \; \|y - W d(\check{\mu}_\star^{\rm nc}) \| .
\end{equation}

\section{Simulation of Artificial Time Series}
\label{section:test_setup}

So far we have considered a general signal which includes a deterministic component and a stochastic component. The parametrisation of each component depends on prior knowledge about the physics of the stellar oscillations. Solar-like pulsations are stochastic in nature and no deterministic component is needed in this case. On the other hand, long-lived stellar pulsations are treated as deterministic. Some stars may support both deterministic and stochastic oscillations.  In this section, we model the two cases separately.

We want to test the fitting method [Equations (\ref{equ:likelihood}) and (\ref{equ:likelihood-estimator})] by applying it to simulated time series with gaps. For comparison, we also want to apply the old fitting method (Section~\ref{section:fitting_nocorr}) to the same time series. We need to generate many realizations of the same random process in order to test the estimators for bias and precision: this is called Monte-Carlo simulation. In Section~\ref{section:window_function}  we discuss the generation of the synthetic window functions. We then discuss the parametrisation of the solar-like oscillations (Section~\ref{section:model_solar_like}) and the deterministic oscillations (Section~\ref{section:model_deterministic}) used to simulate the unconvolved signal.

\subsection{Synthetic Window Functions}
\label{section:window_function}

\begin{figure}[t]
\begin{center}
  \includegraphics[width=0.8\textwidth]{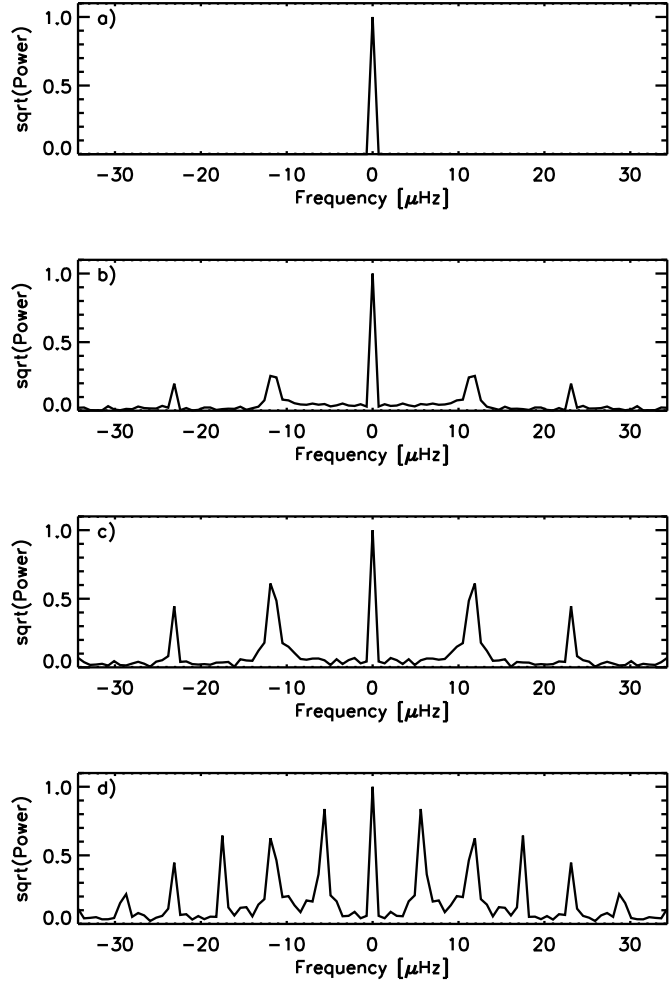}
\end{center}
\caption{
Square root of the power spectra of the synthetic window functions [$\hat{w}$] used in this paper. From top to bottom, the duty cycle is (a) 100\%, (b) 66\%, (c) 30\%,  and (d) 15\%. The main periodicity of the window is 24 hours for cases (b) and (c), and 48~hours for window (d). All windows are truncated at frequency $\nu_p=34.3$~$\mu$Hz.}
\label{figure:windowfunction}
\end{figure} 

We generate three different observation windows, corresponding to different duty cycles.  The observation windows are first constructed in the time domain. By definition, $\tilde{w}_i$ is set to one if an observation is available and zero otherwise. The total length of all time series is fixed at $T=16.5$ days (frequency resolution $\Delta \nu = 0.7\;\mu$Hz). A window function is characterized by two main properties: the duty cycle (fraction of ones) and the average periodicity. A typical window function for a single ground-based site has a 24-hour periodicity. In order to deviate slightly from purely periodic window functions  we introduce some randomness for the end time of each observation block.

Figure~\ref{figure:windowfunction}(b)\,--\,(d) shows the power spectra of the three window functions. The first and second window functions have a main periodicity of 24~hours and duty cycles of $66\%$ and $30\%$ respectively. Two side lobes occur at frequencies $11.6\;\mu$Hz and $23.1\;\mu$Hz. The non-vanishing power between the side lobes is due to the deviation from a purely periodic window.  The third window function has a main periodicity of 48~hours and a duty cycle of only 15\%.  All of these window functions are not unrealistic. 

We apply a sharp low-pass filter at frequency $\nu_p=34.3$~$\mu$Hz ($p=49$) to all window functions. The power at higher frequencies corresponds to about 5\% of the total power in the windows. This truncation is needed to apply the fitting algorithm, which assumes that there exists  a frequency $\nu_p$ beyond which the power in the window vanishes, {\it i.e.} that the window function is band limited.

\subsection{Modeling Solar-Like Oscillations}
\label{section:model_solar_like}

We generate the realizations of the unconvolved solar-like oscillation signal directly in the Fourier domain. We consider a purely stochastic signal ($d=0$) and a single mode of oscillation.  Since we assumed stationarity in the time domain, the Fourier spectrum of the unconvolved signal for one single mode can be written as
\begin{equation}
  x_i = e_i = \left[ \cS L(\nu_i) + \cN \right]^{1/2} \eta_i, \quad i=0,1,\dots,M+2p-1 ,
\end{equation}
where  $L$ describes the line profile of the mode in the power spectrum, $\cS$ is the mode's maximum power, $\cN$ is the variance of the background noise, and $\eta_i$ is a centered complex Gaussian random variable with unit variance and independent real and imaginary parts. Solar-like oscillations are stochastically excited and intrinsically damped by turbulent convection \cite{Goldreich1977,Stein2004}. The expectation value of the power spectrum is nearly Lorentzian, except for some line asymmetry ({\it e.g.}, \opencite{Duvall1993}). Here we use a simple asymmetric line profile:
\begin{equation}
  L(\nu) = \frac{(1 + bX)^2 + b^2}{1+X^2} \quad \mbox{with} \quad X = \frac{\nu - \nu_0}{\Gamma/2} ,
  \label{equ:lorentzian_asymmetric}
\end{equation}
where $\nu_0$ is the resonant frequency, $b$ is the asymmetry parameter of the line profile ($|b|\ll 1$), and $\Gamma$ is a measure of the width of the line profile. We refer to $\cS/\cN$ as the signal-to-noise ratio in the power spectrum. As $b$ tends to zero, $\Gamma$ becomes the full width at half maximum (FWHM) of the power spectrum and $1/(\pi \Gamma)$  the mode lifetime.
There are five model parameters,  $\mu=(\nu_0, \Gamma, b, \cS, \cN)$.

Once the unconvolved signal $x$ has been generated in the Fourier domain, the observed signal $y$ is obtained by multiplication with the window matrix $W$, as explained above.

\subsection{Modeling Deterministic Sinusoidal Oscillations plus White Noise}
\label{section:model_deterministic}

In the time domain, we consider a purely sinusoidal function on top of white background noise:
\begin{equation}
  \tilde{x}_i = A \sin \left( 2\pi \nu_0 \,  t_i + \varphi \right)  + \sigma_t \; {\eta}_i  \qquad i=0,1,\dots,N-1 .
\end{equation}
The first term describes the deterministic component of the signal, where $A$ is the amplitude, $\nu_0$ the mode frequency, and $\varphi$ the phase of the mode.  The second term is stochastic noise with standard deviation $\sigma_t$. The ${\eta}_i$ are $N$ normally distributed independent real random variables with zero mean and unit variance.  The observed signal is obtained by multiplying $\tilde{x}_i$ by the window $\tilde{w}_i$ in the time domain.  The model parameters are $\mu=(\nu_0, \varphi, A, \sigma_t)$.
 
We have defined the signal and the noise in the time domain, but a definition of signal-to-noise ratio in the Fourier domain is desirable.  On the one hand, the variance of the noise in the Fourier domain is
\begin{equation}
\sigma_{\rm n}^2  =  \sigma_0^2 \sum_{i=-p}^{p} |\hat{w}_i|^2 = \frac{\sigma_t^2}{N} \sum_{i=-p}^{p} |\hat{w}_i|^2  ,
\end{equation}
where $\sum_i |\hat{w}_i|^2$ is the total power in the window. On the other hand, the maximum power of the signal in Fourier space is  $P_{\rm max}  =  A^2|\hat{w}_0|^2 / 4$,
where $|\hat{w}_0|^2$ is the power of the window at zero frequency. Thus, by analogy with the solar-like case, it makes sense to define the signal to noise ratio in the Fourier domain as
\begin{equation}
  \cS/\cN  =
  \left( \frac{N |\hat{w}_0|^2}{ 4 \sum_i |\hat{w}_i|^2 }  \right) \frac{A^2}{\sigma_t^2}.
  \label{equ:sn_det}
\end{equation}
In practice we fix $A$ and $\cS/\cN$ and deduce the corresponding noise level $\sigma_t$.

\section{Testing the Fitting Methods}
\label{section:fitting_results}

Several  hundreds of realizations  are needed in order to assess the quality of a fitting method. We do not intend to test all possible combinations of mode parameters but we want to show a few cases for which the new fitting method provides a significant improvement compared to the old fitting method.

\subsection{Solar-Like Oscillations: Window Function with 30\% Duty Cycle}
\label{sec:results_solar_like}

\begin{figure}[t]
  \centering
  \includegraphics[width=\textwidth]{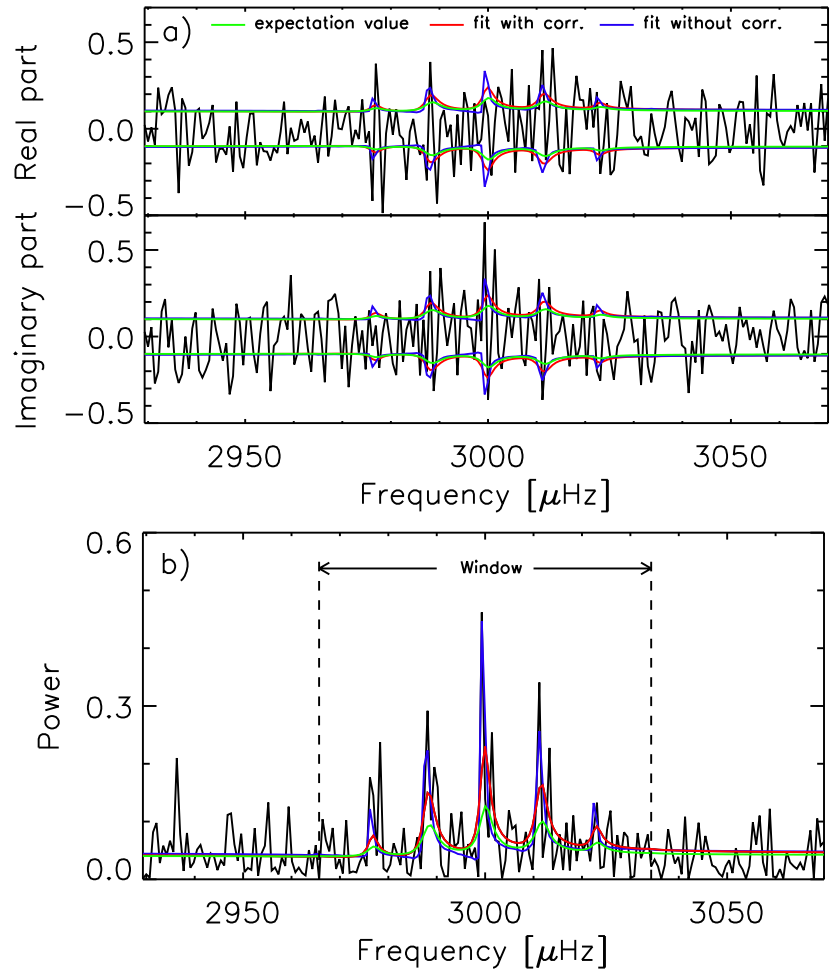}
  \caption{Example of a realization of one mode of a solar-like oscillation (black line) with input frequency $\nu_0 = 3000\;\mu$Hz,  linewidth $\Gamma = 3.2\;\mu$Hz, and $\cS/\cN=6$. The window function is 30\% full. Panel (a) shows the real and imaginary parts of the Fourier spectrum. Panel (b) shows the power spectrum. The vertical dashed lines represent the width of the window function. Also shown are the new fit (red), the old fit (blue), and the expectation value (green). }
  \label{fig:realization_fit_stoch}
\end{figure}

Figure~\ref{fig:realization_fit_stoch} shows one realization of a simulated  mode of solar-like oscillation with input parameters  $\nu_0=3000$~$\mu$Hz, 
$\Gamma=3.2$~$\mu$Hz, $\cS=0.9$, $\cN=0.15$, and $b=0.1$. The signal to noise ratio is $\cS/\cN=6$ 
and the window function is 30\% full (see Figure~\ref{figure:windowfunction}(c)).
The mode lifetime is $1/(\pi\Gamma)=27.6$~hours.
 Figure~\ref{fig:realization_fit_stoch}(a) displays the real and imaginary parts
of the Fourier transform $y$, together with the standard deviation of the data (nc fit in blue, new fit in red, expectation value in green). 
Figure \ref{fig:realization_fit_stoch}(b) shows the power spectrum and the fits. Notice the sidelobes introduced by the convolution of the signal with
the window functions.
The ``no-correlation'' fit is done on the power spectrum [Equation~(\ref{eq.nc})], while the new fit is performed in complex Fourier space [Equation~(\ref{equ:likelihood})].

Each fit shown in Figure~\ref{fig:realization_fit_stoch} corresponds in fact to the best fit out of five fits with different initial guesses. For each realization, we use the frequency guesses  $3000+(0, \pm 5.5, \pm 11.9)$~$\mu$Hz for $\nu_0$. The last two frequency guesses correspond to the frequencies of the two main sidelobes of the window function (Figure~\ref{figure:windowfunction}c). For the other parameters,
we choose random guesses within $\pm20$\% of the input values.  
The reason for using several guesses is to  ensure that the fit converges to the global maximum of the likelihood, not to a nearby local maximum, {\it i.e.} that the estimates returned by the code are the MLE estimates defined by Equation~(\ref{equ:likelihood-estimator}). In some cases, the global maximum coincides with a sidelobe at $\pm 11.9$~$\mu$Hz from the main peak. We note that the new fitting method requires a much longer computing time than the old nc method: typically,  three hours on a single CPU core for a single realization (five guesses, five fits).

For the particular realization of Figure~\ref{fig:realization_fit_stoch}, the new fit is closer to the expectation value ({\it i.e.} is closer to the answer) 
than the old nc fit.  No conclusions should be drawn, however, from looking at a single realization.  

\begin{figure}[t]
  \centering
  \includegraphics[width=\textwidth]{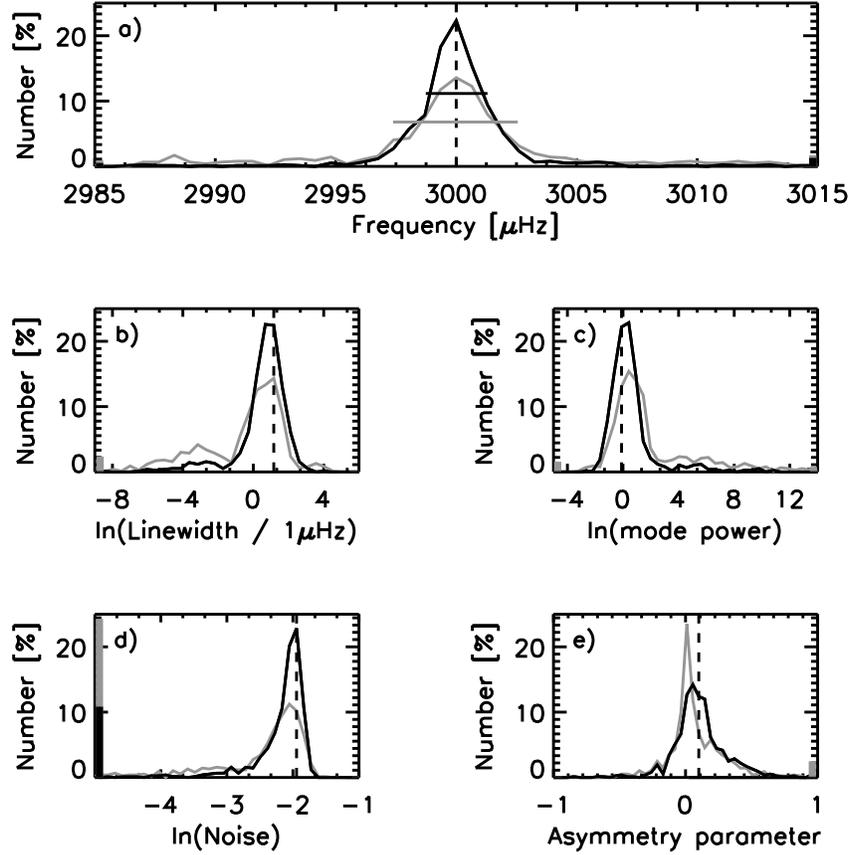}
  \caption{Distributions of the inferred oscillation parameters from fits to 750 realizations of a single mode of solar-like oscillation. The input parameters are given in Table~1 and the window function is 30\% full. The five panels show the distributions of the inferred (a) mode frequency $\nu_0$, (b) linewidth $\Gamma$, (c) mode power $\cS$, (d) noise level $\cN$, and (e) asymmetry parameter $b$. The black lines show the results obtained with the new fitting method and the grey lines show the old ``no-correlation'' fits. The vertical dashed line in each plot indicates the input value. The horizontal lines in panel (a) are intervals containing 68\% of the fits for the new (black line) and the old (grey line) fitting methods. The thick black and grey vertical lines in panel (d) give the numbers of outliers with $\ln N < -5$.}
  \label{fig:distr_stoch_sn30_wa}
\end{figure}

In order to test the reliability of each fitting method, we computed a total of 750 realizations 
with the same input parameters  as in Figure~\ref{fig:realization_fit_stoch} and the same window function (30\% full).  
The quality (bias and precision) of the estimators can be studied from the distributions of the inferred parameters.
As shown by the distributions of Figure~\ref{fig:distr_stoch_sn30_wa} the new fitting method is superior to the old nc method.
This is true for all the parameters, in particular the mode frequency $\nu_0$.
The distributions for the mode frequency (Figure~\ref{fig:distr_stoch_sn30_wa}(a)) are quite symmetric and Gaussian-like, although the old fitting method leads to a significant excess of values beyond the two-$\sigma$  mark.
We note that, in general, the old fitting method is more sensitive to the initial frequency guess. 
Also the estimates of the linewidth $\Gamma$ and the mode power $\cS$ are significantly more biased with the old fitting method than with the new one
(Figures~\ref{fig:distr_stoch_sn30_wa}b, \ref{fig:distr_stoch_sn30_wa}c). 
It is worth noting that the fits return a number of small $\Gamma$/large $\cS$ estimates away from the main peaks of the distributions, less so for the new fits. These values correspond to instances when the signal barely comes out of the noise background.
The new fit returns the noise level [$\cN$] with a higher precision and a lower number of underestimated outliers than the old method (the ouliers are represented by the vertical bars in Figure~\ref{fig:distr_stoch_sn30_wa}(d)).
Although the estimation of the asymmetry parameter is unbiased with the new fitting method (Figure~\ref{fig:distr_stoch_sn30_wa}(e)), the uncertainty on $b$ is so large that it probably could have been ignored in the model. 

Quantitative estimates of the mean and the dispersion of the estimators are provided in Table~1. Because the 
distributions of the estimated parameters are not always Gaussian and may contain several outliers, 
we compute the median (instead of the mean) 
and the lower and upper bounds corresponding to $\pm 34$\% of the points on each side of the median (instead of the one-$\sigma$ dispersion).  This definition has the advantage of being robust with respect to the outliers. The notation $3000.0^{+b}_{-a}$~$\mu$Hz in the first row of Table~\ref{table:freq_uncertainty} means that 
the median mode frequency is $3000.0$~$\mu$Hz and that $68\%$ of the fits belong to the interval $[3000.0-a, 3000.0+b]$~$\mu$Hz. We emphasize that the subscript $-a$ and the superscript $+b$ do not refer to an uncertainty in the determination of the median: the median is known to a much higher precision thanks to the large number of realizations. Later we relax the language and refer to the ``one-$\sigma$ uncertainty'' to mean the average  $\sigma=(a+b)/2$.

\begin{table}[t]
  \caption{Medians and scatters of the distributions of the estimated parameters of solar-like oscillation (see Figure~\ref{fig:distr_stoch_sn30_wa}). The window function is 30\% full, the input linewidth is $3.2$~$\mu$Hz, and the input signal-to-noise ratio is ${\cal S}/{\cal N}=6$. The new and old MLE estimates are given in the last two columns. By definition, 68\% of the fits fall within the bounds set by the subscripts/superscripts (the notation is explained in detail in the text). }
  \label{table:freq_uncertainty}
  \begin{tabular}{rrrr}
    \hline Mode parameter
    & Input value & New fitting & Old fitting \\
    \hline\\[-5pt]
    $\nu_0$ [$\mu$Hz] & $3000.0$ & $3000.0 {+1.4 \atop -1.4}$ & $3000.0 {+2.8 \atop -2.8}$\\[10pt]
    $\ln (\Gamma [\mu{\rm Hz}])$ & $1.2$ & $0.8 {+0.8 \atop -1.0}$ & $0.2 {+1.1 \atop -3.7}$\\[10pt]
    $\ln \cS$ & $-0.1$ & $0.2 {+0.9 \atop -0.9}$ & $0.9 {+4.3 \atop -1.2}$\\[10pt]
$\ln \cN$ & $-1.9$ & $-2.1 {+0.2 \atop -0.9}$ & $-2.4 {+0.4 \atop -6.8}$ \\[10pt]
    $b$ & $0.1$ & $0.1 {+0.2 \atop -0.1}$ & $0.0 {+0.2 \atop -0.1}$\\[5pt]
    \hline
  \end{tabular}
\end{table}

The numbers from the last two columns in Table~1 confirm the analysis of Figure~\ref{fig:distr_stoch_sn30_wa}. The mode frequency can be measured with a precision of $1.4$~$\mu$Hz, which is exactly twice better with the new fitting method than with the old one. 
 This gain in precision is very significant and potentially important. Since measurement uncertainty scales like $T^{-1/2}$ \cite{Libbrecht1992}, one may equate the gain in using the proper fitting procedure to an effective increase in the total length of the time series  by a factor of four. As seen in Table~1, the  linewidth, the mode power, the background noise, and the line asymmetry parameter are all less biased and more precise with the new fitting method than the old one. 
Notice that the larger dispersions in the old-fit case are due in part to non-Gaussian distributions with extended tails.

\begin{figure}[t]
  \includegraphics[angle=90,width=\textwidth]{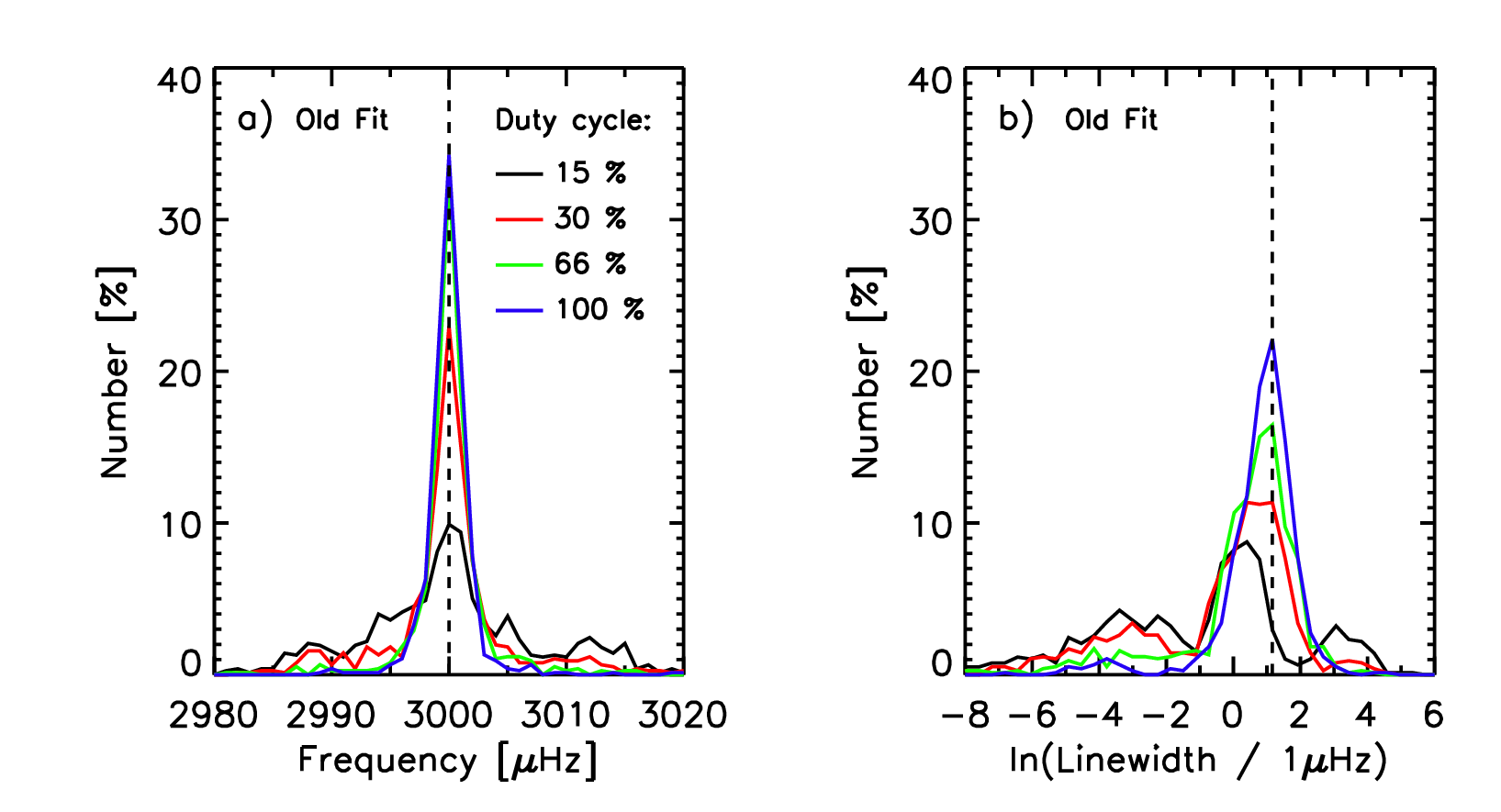}\\
  \includegraphics[angle=90,width=\textwidth]{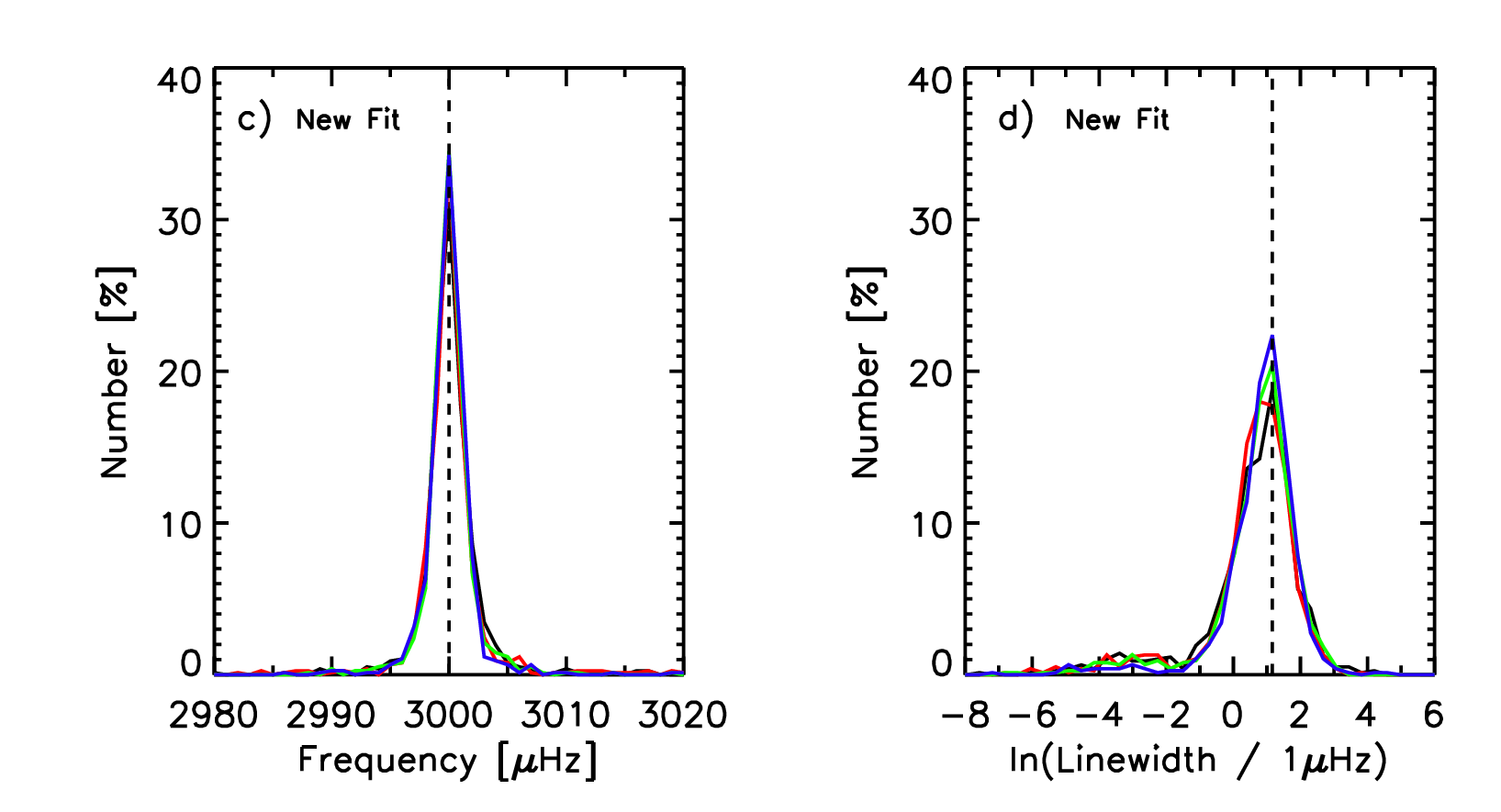}
  \caption{Distributions of the mode frequency and the linewidth for 750 realizations of solar-like oscillations, using the old fitting method (panels a and b) and the new fitting method (panels c and d). The observation windows have a duty cycle of 15\%, 30\%, 66\%, and 100\%. The vertical dashed lines represent the input values. The input linewidth is $\Gamma=3.2$~$\mu$Hz}
  \label{fig:distribution_window}
\end{figure}

\begin{table}[t]
  \caption{Medians and scatters of the mode frequency estimates (solar-like oscillations) for the window functions defined in Section~\ref{section:window_function}. The input mode frequency is $\nu_0 = 3000\;\mu$Hz, the input linewidth is $\Gamma = 3.2\;\mu$Hz, and the signal-to-noise ratio is fixed at $\cS/\cN=6$. The mode lifetime is $27.6$~hours.}
  \label{table:freq_uncertainty2}
  \begin{tabular}{cccccc}
    \hline
    \multicolumn{3}{c}{Window function} & \multicolumn{2}{c}{Frequency estimate [$\mu$Hz]} \\
    Duty cycle  & Main period & Average gap & New fitting & Old fitting\\
    \hline\\[-5pt]
    100\% & -- & -- & $ 3000.0 {+1.1 \atop -1.2}$ & $ 3000.0 {+1.1 \atop -1.2}$ \\[10pt]
    66\% & 24~hours & 7.4~hours & $ 3000.0 {+1.1 \atop -1.3}$ & $ 3000.1 {+1.5 \atop -1.4}$ \\[10pt]
    30\% & 24~hours & 16.4~hours & $ 3000.0 {+1.4 \atop -1.4}$ & $ 3000.0 {+2.8 \atop -2.8}$ \\[10pt]
    15\% & 48~hours & 40.7~hours & $ 3000.0 {+1.7 \atop -1.3}$ & $ 3000.0 {+8.3 \atop -6.5}$ \\[5pt]
    \hline
  \end{tabular}
\end{table}

\begin{figure}[h]
  \includegraphics[width=\textwidth]{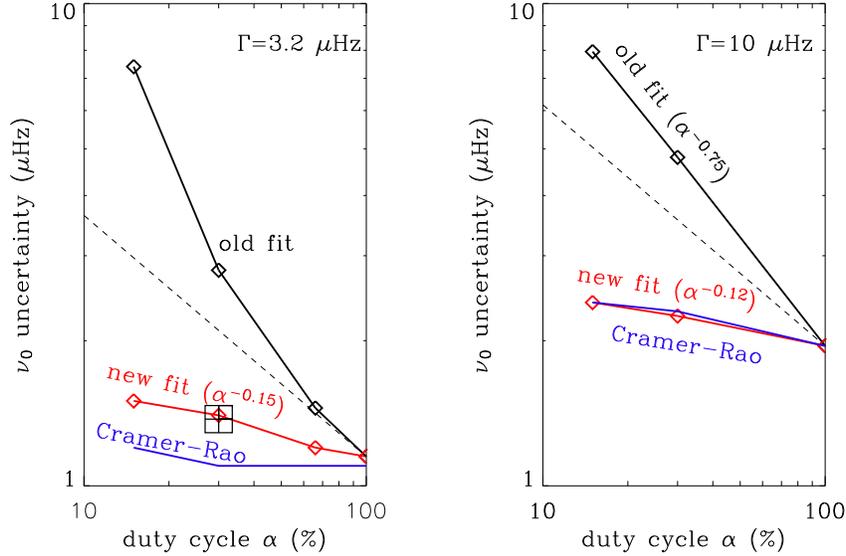}
  \caption{Uncertainty on estimates of the mode frequency [$\nu_0$] as a function of the window duty cycle [$\alpha$]. The window functions are as defined in Section~\ref{section:window_function}.  The red curve shows the 1-$\sigma$ Monte-Carlo MLE uncertainties for the new fitting method. The black curve shows the 1-$\sigma$ Monte-Carlo MLE uncertainties for the old no-correlation fitting method. The blue curves show the mean Cram\'er-Rao lower bounds (formal error bars).
The square symbol with a cross at $\alpha=30\%$ in the left panel is a rough estimate (see text).  In the left panel, the input linewidth is $\Gamma=3.2$~$\mu$Hz (see also numbers in Table 1). In the right panel, the input linewidth is $\Gamma=10$~$\mu$Hz, all of the other parameters being the same as in the left panel. In both panels the signal-to-noise ratio is $\cS/\cN=6$.
For reference, the dashed lines have slope $\alpha^{-1/2}$. }
  \label{fig:sigma_nu0}
\end{figure}

\subsection{Solar-Like Oscillations: Different Window Functions}
Here we study how bias and precision change as the window function changes, in particular as the duty cycle changes. We consider the four window functions defined in section \ref{section:window_function} with   
  duty cycles [$\alpha$] equal to 15\%, 30\%, 66\%, and 100\%. 
First we consider input parameters of solar-like oscillations that are exactly the same as in the previous section: $\nu_0=3000$~$\mu$Hz, $\Gamma=3.2$~$\mu$Hz, $\cS=0.9$, $\cS/\cN=6$, and $b=0.1$.
Figure~\ref{fig:distribution_window} shows the distributions of the inferred mode frequencies and linewidths, using the old (Figures~\ref{fig:distribution_window}a, \ref{fig:distribution_window}b) and the new  (Figures~\ref{fig:distribution_window}(c), \ref{fig:distribution_window}(d)) fitting methods. Each fit is the best fit from five different $\nu_0$ guesses  (see Section \ref{sec:results_solar_like}). The distributions for the 100\%-window are identical for the two fitting methods; this is expected since the old and new fitting methods are equivalent in the absence of gaps. 

The precision on $\nu_0$ using the old ``no-correlation'' MLE drops fast as the duty cycle decreases (Figure~\ref{fig:distribution_window}(a)). This drop is much faster than in the case of the fits that take the frequency correlations into account (Figure~\ref{fig:distribution_window}(c)). When the duty cycle is $15\%$, the frequency estimate is five times better with the new than the old method.
The difference is perhaps even more obvious for the linewidth. For the 15\% window, it is almost impossible to retrieve  $\Gamma$  with the old fitting method (Figure \ref{fig:distribution_window}b), while the new method gives estimates that are almost as precise as in the no-gap case (Figure~\ref{fig:distribution_window}d). The estimates of $\Gamma$ are significantly less biased with the new method.
Figure~\ref{fig:distribution_window} confirms the importance of using the correct expression for the likelihood function.

Table~\ref{table:freq_uncertainty2} gives the medians and half-widths of the $\nu_0$ distributions. The one-$\sigma$ dispersions are plotted as a function of the duty cycle $\alpha$ in the left panel of Figure~\ref{fig:sigma_nu0}. The improvement in the fits is quite spectacular. For example, when $\alpha=15\%$ the dispersion on $\nu_0$ is five times less with the new fitting method ($1.5$~$\mu$Hz vs. $7.4$~$\mu$Hz).

With the old method, the uncertainty on $\nu_0$ increases much faster than $\alpha^{-1/2}$ as the duty cycle $\alpha$ drops ($\sim \alpha^{-1}$ between the 30\% and 15\% windows). This steep dependence on $\alpha$ is worse than ``predicted'' by \inlinecite{Libbrecht1992}. In his paper, Libbrecht suggested to use the uncertainty $\sigma_{\nu_0} = \sqrt{f  \Gamma / (4\pi T)}$ where $f(\beta)=(1+\beta)^{1/2}[(1+\beta)^{1/2}+\beta^{1/2}]^3$ and $\beta$ is an ``effective'' noise-to-signal ratio. He suggested that the main effect  of the gaps is to increase the noise-to-signal ratio ${\cal N}/{\cal S}$, presumably by a factor $\sum_i |\hat{w}_i|^2 / |\hat{w}_0|^2$, itself proportional to $1/\alpha$. This leads however to a dependence of $\sigma_{\nu_0}$ on $\alpha$ which, in our particular case, is closer to $\alpha^{-1/2}$ than $\alpha^{-1}$. We suspect that the Libbrecht formula underestimates the dispersion because it ignores the frequency correlations.  

The new fitting method returns a $\nu_0$ uncertainty that is much less sensitive to the duty cycle, with a variation like $\sim\alpha^{-0.15}$ (red curve, left panel of Figure~\ref{fig:sigma_nu0}). This is quite remarkable. That the frequency uncertainty could remain nearly constant for $\alpha>30\%$ is not really surprising since the average gap (see numbers in Table~2) is less than the mode lifetime $\tau = 1/(\pi \Gamma) = 27.6$~hours. This regime was studied by \inlinecite{Fossat1999} using a gap-filling method: as long as the signal-to-noise ratio is large enough, the signal can be reconstructed.  Why the new fit is doing such a good job for duty cycles $\alpha \leq 30\%$ is, however,  puzzling (at first sight), since the average gap ($40.7$~hours) is larger than the mode lifetime.  This can be understood as follows. For small duty cycles, the time-series is effectively a collection of nearly independent blocks of data, which, for the 30\% window function, are eight-hour long on average. Since MLE simulations tell us that the uncertainty on the mode frequency for an uninterrupted series of eight hours is about $5.5$~$\mu$Hz, we would expect for the gapped time series ($T=16.5$~days, 24-hour periodicity) to be able to reach the uncertainty $5.5/\sqrt{16}=1.375$~$\mu$Hz.  This value, represented by the box with a cross in Figure~6, is found to be very close to the MLE estimate from the new fits. Hence, what matters at very low duty cycle is the number of independent blocks of continuous data. The new fitting method captures this very well, which is satisfying. By comparison, the old no-correlation fitting method does poorly (black line). 

In order to further investigate this last point, we ran another set of simulations using a mode linewidth $\Gamma = 10\;\mu$Hz corresponding to a mode lifetime $\tau = 8.8$~hours, which is significantly smaller than the average gap lengths of the 30\% and the 15\% windows.  The other input parameters remained the same as above. We computed and fitted  1350 realizations. The results are shown in the right panel of Figure~6. For the new fitting method, the dependence of the 
frequency uncertainty on the duty cycle is about $\alpha^{-0.12}$, which is comparable to the previous simulations with $\Gamma=3.2$~$\mu$Hz. 
We conclude that it is really worth solving for the correct minimization problem and that fitting for the phase information in complex Fourier space is important to get a good match between the model and the data. Of course, this can only be done properly when we have a perfect knowledge of the model, which is the case with these numerical simulations, but is rarely the case with real observations.

\begin{figure}[t]
  \includegraphics[angle=90,width=\textwidth]{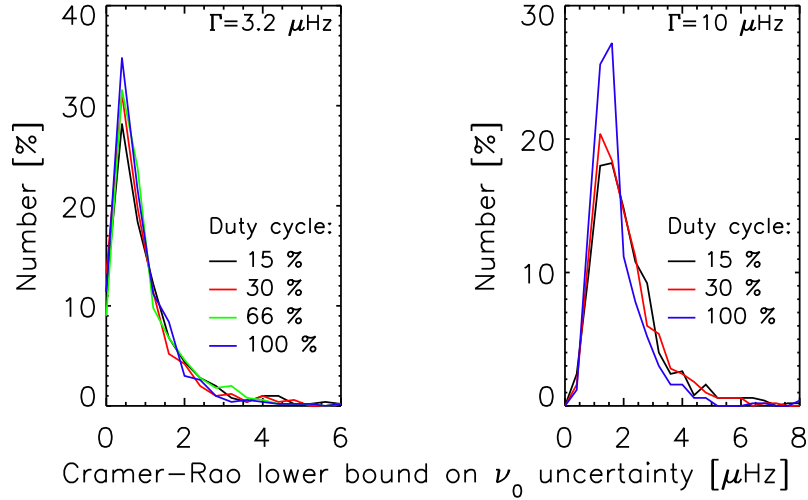}
  \caption{Distributions of formal errors on the mode frequency obtained by inverting the Hessian. The left panel is for the simulation with $\Gamma=3.2$~$\mu$Hz (see Figure 5c) and the right panel is for $\Gamma=10$~$\mu$Hz. The different curves correspond to different window functions, as indicated in the legend. The means of these distributions (Cram\'er-Rao lower bounds) give the blue curves plotted in Figure~\ref{fig:sigma_nu0}.}
  \label{fig:cramer}
\end{figure}

\subsection{Solar-Like Oscillations: Cram\'er-Rao Lower Bounds}

Monte-Carlo simulations are very useful in order to assess the variance and the bias of a particular estimator. When fitting real observations, however, the variance of the estimator cannot be computed directly by Monte-Carlo simulation since the input parameters are, by definition, not known. Hopefully, the fit can return a formal error from the shape of the likelihood function in the neighborhood of the global maximum. 

The Cram\'er-Rao lower bound \cite{Kendall1967} achieves minimum variance among unbiased estimators. It is obtained by expanding ${\cal L}$ about its minimum. The formal error $\sigma_{\mu_i}$ on the parameter $\mu_i$ is given by
\begin{equation}
\sigma_{\mu_i} = \sqrt{ K_{ii} } ,
\end{equation}
where $K_{ii}$ is the $i$th element on the diagonal of the inverse [$K=H^{-1}$]  of the Hessian matrix with elements
\begin{equation}
H_{i j}  = \frac{\partial^2 {\cal L} }{\partial \mu_i \; \partial \mu_j }  
\quad {\rm for} \quad i,j = 0,1,\dots,k-1
.
\end{equation}
The Cram\'er-Rao formal errors have been used in helioseismology by, for example,  \inlinecite{Toutain1994}, \inlinecite{Appourchaux1998}, and \inlinecite{Gizon2003}.

We have computed the formal error on the mode frequency for many realizations and for all window functions. The resulting distributions are shown in Figure~7. The mean formal error from each distribution is plotted in Figure~6.  Overall the Cram\'er-Rao lower bound is remarkably close to the Monte-Carlo MLE uncertainty using the new fitting method; they are even undistinguishable when $\Gamma=10$~$\mu$Hz.

This is useful information as it means that, on average, the Hessian method provides reasonable error estimates. It should be clear, however, that the distributions shown in Figure~7 show a significant amount of scatter: the formal error from the Hessian may be misleading for particular realizations.

\subsection{Sinusoidal Deterministic Oscillation plus White Noise}
\label{section:results_deterministic}

\begin{figure}[t]
  \centering
  \includegraphics[width=\textwidth]{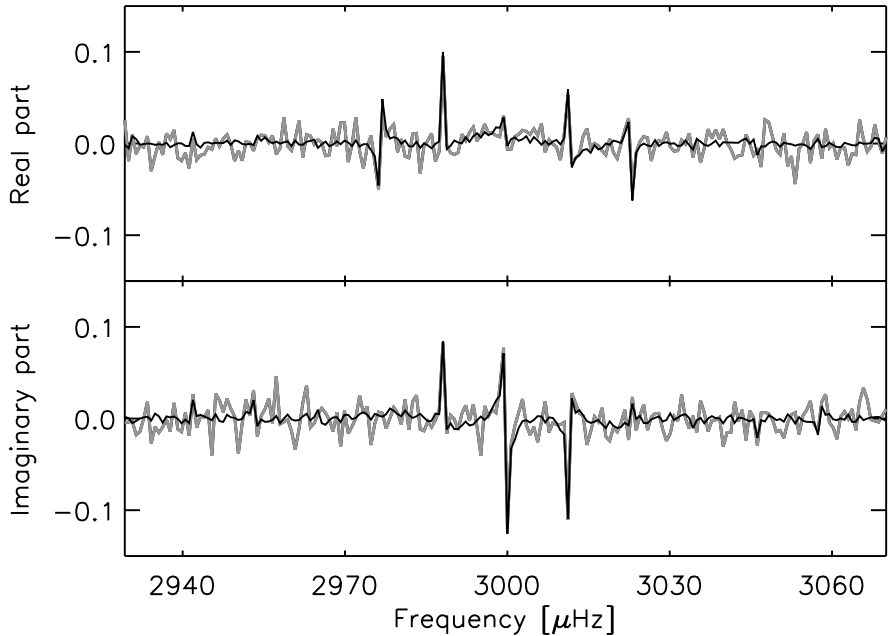}
  \caption{Real and imaginary part of the Fourier transform of a simulated gapped time series containing a sinusoid on top of  white background noise. The signal-to-noise ratio is $\cS/\cN=100$. The observation window has a duty cycle of 30\%. The simulated data is the thick grey line. The thin black line shows the fit to the data using the new fitting method. The fit with the old method is not shown since it is almost identical. }
  \label{fig:realization_fit_determin}
\end{figure}

Figure \ref{fig:realization_fit_determin} shows the Fourier spectrum of a simulated time series containing a sinusoidal mode of oscillation on top of a white noise background as described in Section \ref{section:model_deterministic}. In this particular case the observation window with a duty cycle of 30\% is used (see Figure~\ref{figure:windowfunction}(c)). The input parameters of the sinusoidal function are the mode frequency $\nu_0=3000\;\mu$Hz, the amplitude $A=1.1$, and the phase $\varphi = 60^\circ$. The signal-to-noise ratio is ${\cal S}/{\cal N}=100$. The fit shown in Figure \ref{fig:realization_fit_determin} was obtained with the new fitting method. Since we found no significant difference between the old and the new fitting methods in this case, the old fitting method is not shown. Differences between the data and the fit are essentially due to the noise.

\begin{figure}[t]
  \centering
  \includegraphics[angle=90,width=\textwidth]{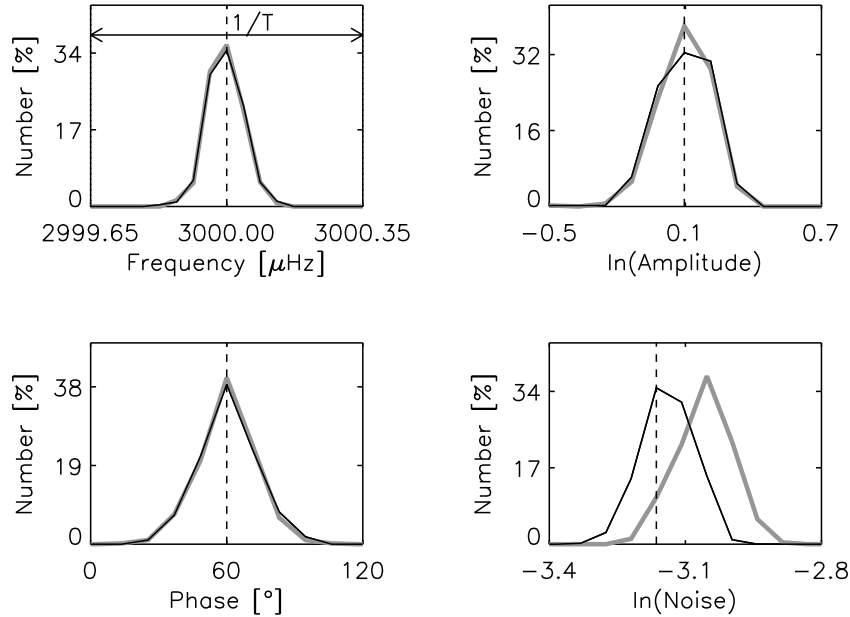}
  \caption{Distributions of the inferred oscillation parameters for a set of 500 realizations of long-lived sinusoidal oscillations with $\cS/\cN=46$. The window function with a duty cycle of 30\% is used. The black and the grey lines are for  the new and old fitting methods respectively. The vertical dashed line in each plot indicates the input value. The parameters shown are (a) the mode frequency [$\nu_0$], (b) the logarithm of the mode amplitude [$\ln A$], (c) the phase of the oscillation  [$\phi$], and (d) the logarithm of the noise level [$\ln \sigma_0$] (see Section~6.3). Notice that the estimate of the noise is biased when frequency correlations are ignored (old nc fit), although by a very  small amount.}
  \label{fig:distribution_deterministic}
\end{figure}

We computed 500 realizations of sinusoidal oscillations with the same mode parameters (frequency, amplitude, phase) as above, the same observation window (30\% full), but with a signal-to-noise ratio $\cS/\cN=46$. The resulting distributions of the inferred parameters obtained with the two fitting methods are shown in Figure \ref{fig:distribution_deterministic}. For this simulation, the known input values were used as an initial guess to speed up the minimization; we checked on several realizations that it is acceptable to do so when the signal to noise ratio is large. The distributions of the inferred parameters (Figure \ref{fig:distribution_deterministic}) show that for sinusoidal oscillations, the new fitting method does not provide any significant improvement compared to the old fitting method. 

\begin{figure}[t]
  \centering
  \includegraphics[angle=90,width=0.9\textwidth]{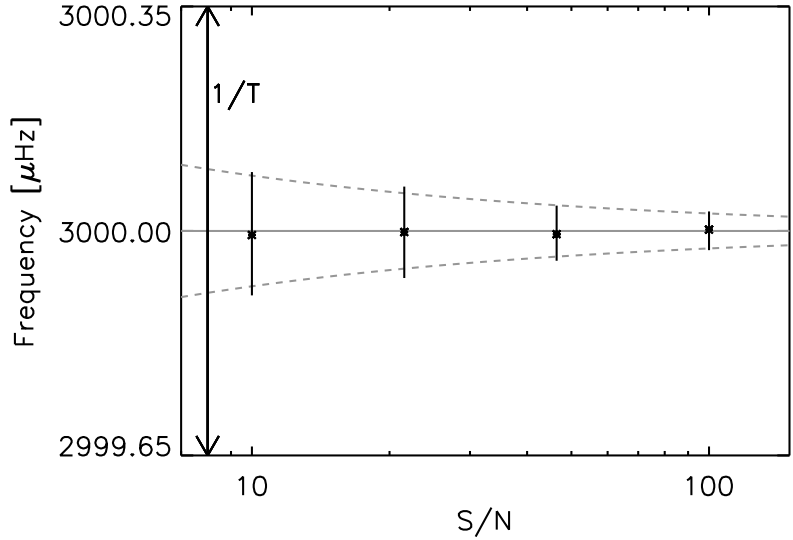}
  \caption{Median (cross) and standard deviation (vertical bar) of the inferred frequency of sinusoidal oscillation [$\nu_0$] as a function of signal-to-noise ratio $\cS/\cN$.  The duty cycle is 30\,\%. Only the results obtained with the new fitting method are shown. The horizontal grey line shows the input mode frequency. The dashed grey lines show the theoretical value of frequency uncertainty, $\sigma_{\nu_0}$, given by Equation~(\ref{equ:freq_error_determin}). The vertical axis of the plot spans the interval $\Delta \nu = 1/T = 0.7\;\mu$Hz.}
  \label{fig:fsn_det}
\end{figure}

We emphasize that the fitting parameters can be determined with a very high precision when the noise level is small. In particular, we confirm that the uncertainty of the frequency estimator can be  much smaller than $1/T$ (see Figure \ref{fig:distribution_deterministic}(a)).  Figure \ref{fig:fsn_det} shows the median and the standard deviation of the mode frequency for different signal-to-noise ratios. Each symbol and its error bar in Figure \ref{fig:fsn_det} is based on the computation of 500 realizations of sinusoidal oscillations with the same mode parameters as above, the same observation window (30\% full), but various signal-to-noise ratios. Since we did not find any significant difference between the two fitting methods, only the results obtained with the new fitting method are shown. Figure \ref{fig:fsn_det} illustrates that even for a relatively low signal-to-noise ratio of $\cS/\cN=10$, the standard deviation of the inferred mode frequency is smaller than $1/T$ by a factor of four. For higher signal-to-noise ratios the precision is even more impressive: when  $\cS/\cN=100$, the standard deviation of the mode frequency is about 20 times smaller than $1/T$.

The theoretical value of the standard deviation of the mode frequency obtained by \inlinecite{Cuypers1987} can be extended to the case of gapped data (Cuypers, 2008, private communication) as follows:
\begin{equation}
  \sigma_{\nu_0} = \frac{\sqrt{6} \;  \sigma_{\rm t}}{\pi A T \sqrt{n} }, 
  \label{equ:freq_error_determin}
\end{equation}
where $A$ is the amplitude of the sinusoid in the time domain, $\sigma_{\rm t}$ is the rms value of the noise, $n=\alpha N$ is the number of recorded data points, and $T$ is the total observation length. This theoretical uncertainty is overplotted in Figure \ref{fig:fsn_det}. The match with our Monte-carlo measurements is excellent. This confirms that, in this case, it is equivalent to perform the  fits in the temporal and in the Fourier domains. Note that Equation (\ref{equ:freq_error_determin}) is only valid under the assumption that the noise is uncorrelated in the time domain, a condition fulfilled by our simulations.
The main reason why the measurement precision is only limited by the noise-to-signal ratio is because perfect knowledge of the model is assumed.

\section{Conclusion}
\label{section:conclusions}

In this paper we derived an expression for the joint PDF of solar or stellar oscillations in complex Fourier space, in agreement with the work of \inlinecite{Gabriel1994}. This joint PDF explicitly takes into account frequency correlations introduced by the convolution with the window function. We implemented a maximum likelihood estimation method to retrieve the parameters of stellar oscillations. Both stochastic solar-like oscillations and deterministic sinusoidal oscillations were considered.

In the case of solar-like oscillations, we performed Monte-Carlo simulations to show that the improvement provided by our fitting method can be very significant in comparison with a fitting method that ignores the frequency correlations. The results are summarized in Figure~6. In one particular example, using an observation window with a duty cycle $\alpha=30$\,\% and a signal-to-noise ratio $\cS/\cN=6$, the new fitting method increased the precision of the mode frequency by a factor of two and the estimates of the linewidth and mode power were less biased and more precise. For a window with a duty cycle $\alpha=15$\,\%, the precision on the mode frequency estimate was increased by a factor of five.
We also found that the Cram\'er-Rao lower bounds (formal errors) can provide reasonable estimates of the uncertainty on the MLE estimates of the oscillation parameters.

In the case of long-lived, purely sinusoidal oscillations, we did not find any significant improvement in using this new fitting method. Yet, we confirm that the standard deviation of the mode frequency can be measured in Fourier space with a precision much better than $1/T$ for large signal-to-noise ratios, in accordance with a previous time-domain calculation (Cuypers, 1987; Cuypers, 2008, private communication).

The analysis of time series containing many gaps can benefit from our work.   Applications may include, for example, the re-analysis of solar oscillations from the early days of the BiSON network \cite{Miller2004} or  the solar-like oscillations of $\alpha$ Centauri observed from the ground with two telescopes \cite{Butler2004}.

\begin{acks}
  We thank T.~Appourchaux for useful discussions, in particular for the suggestion to compute the Cram\'er-Rao lower bounds. T.~Stahn is a member of  the International Max Planck Research School on Physical Processes in the Solar System and Beyond at the Universities of G\"ottingen and Braunschweig.  The MLE source code is available from the internet platform of the European Helio- and Asteroseismology Network (HELAS, funded by the European Union) at \url{http://www.mps.mpg.de/projects/seismo/MLE_SoftwarePackage/}.
\end{acks}


\end{article} 

\end{document}